\documentclass[aps,prl,twocolumn,showpacs]{revtex4}

\usepackage{graphicx}

\begin{document}

\preprint{Paper submitted to ...}

\title{The Reconstruction of Pt(111) and \\ Domain Patterns on Close-packed 
Metal Surfaces}

\author{Shobhana Narasimhan}
\email{ shobhana@jncasr.ac.in}
\author{Raghani Pushpa} 
\affiliation{ Theoretical Sciences Unit,
Jawaharlal Nehru Centre for Advanced Scientific Research,\\ Jakkur PO,
Bangalore 560 064, India}


\begin{abstract}

We have studied the reconstruction of Pt(111) theoretically using a
two-dimensional Frenkel-Kontorova model for which all parameters have
been obtained from {\em ab initio} calculations. We find that the
unreconstructed surface  lies right at the stability boundary, and thus
it is relatively easy to induce the surface to reconstruct into a
pattern of FCC and HCP domains, as has been shown experimentally. The
top layer is very slightly rotated relative to the substrate, resulting
in the formation of ``rotors" at intersections of domain walls.  The
size and shape of domains is very sensitive to the density in the top
layer, the chemical potential, and the angle of rotation, with a smooth
and continuous transition from the honeycomb pattern to a Moir\'e
pattern, via interlocking triangles and bright stars. Our results show
clearly that the domain patterns found on several close-packed metal
surfaces are related and topologically equivalent.

\end{abstract}

\pacs{68.35.Bs, 61.72.bb, 68.35.Md, 68.55.-a}

\maketitle


Due to its enormous importance as a catalyst, Pt(111) is one of the
most widely studied surfaces.  Under ``normal" conditions, it has the
flat topography expected of a bulk-truncated face-centered-cubic (FCC)
(111) surface. However, experiments have shown that one can induce the
surface to reconstruct into either a honeycomb structure or a pattern
of interlocking triangles -- for example, by  heating the surface above
1330 K \cite{sandy}, or by placing it in a supersaturated Pt
vapor \cite{bott}. The reconstructed structure is comprised of domains
where the bulk FCC stacking sequence is retained, alternating with
domains where the surface atoms instead occupy hexagonal-close-packed
(HCP) sites, sitting directly above atoms two layers below.

This is one of a family of similar reconstructions, formed by a
tessellation of FCC and HCP domains, seen on Au(111) as well as various
heteroepitaxial systems on the (111) faces of FCC metals, and the
structurally similar (0001) faces of HCP metals
\cite{naau,agpt,curu,niru}.  These structures have attracted a great
deal of interest, especially because  they can be used as templates for
growing ordered arrays of nanoparticles \cite{chambliss,voigt,brune}.
Possible applications for such nanostructures include nanoelectronics,
information storage, and nanoscale chemical reactors.  To design such
nanostructures, it is desirable to understand the factors controlling
the geometry and spacing of the reconstruction patterns.

In this paper, we study the structure of the Pt(111) surface
theoretically. We show that the unreconstructed Pt(111) surface is in
fact teetering right at the brink of a domain of stability.  Thus,
slight changes in the environment can trigger a reconstruction, whose
periodicity and geometry are very sensitive to various extrinsic and
intrinsic parameters. In addition to obtaining excellent agreement with
the structures reported experimentally for Pt(111), we also observe
most of the structures reported experimentally for other systems.

The presence or absence of such reconstructions involves a very
delicate balance between various contributions to the total energy, and
it is therefore desirable to have as accurate a description of
interatomic interactions as possible. Unfortunately, the very large
unit cells of the reconstructed surfaces make it unfeasible to perform
a fully {\em ab initio} calculation.  However, one can use {\em ab
initio} calculations to parametrize a model; this is the approach we
will follow.

The driving force for the reconstruction of these close-packed surfaces
is tensile surface stress, i.e., surface atoms would like to be closer
together than the bulk nearest-neighbor (NN) spacing $a$ \footnote{For
heteroepitaxial systems, the driving force is sometimes compressive
(not tensile) surface stress.}. This tendency is however opposed by the
fact that (in general) it costs energy  when: (i) surface atoms lose registry with
the substrate (ii) extra atoms must be provided to increase the surface
density.  These factors are incorporated in the Frenkel-Kontorova (FK)
model, which has been widely used to study commensurate-incommensurate
transitions \cite{fk}. In its original form, the model consists of a
one-dimensional chain of atoms connected by harmonic springs of
equilibrium length $b \ne a$, sitting in a sinusoidal substrate
potential. This has the advantage of being exactly solvable, and can
serve as a useful guideline for studying structural stability.

However, a proper description requires a generalized version of the FK
model, with a two-dimensional (2D) layer of atoms interacting via a
more realistic anharmonic potential $V_{ss}$, and sitting in a 2D
potential due to bulk atoms, $V_{sb}$, with competing minima at FCC and
HCP sites. The Hamiltonian is given by \cite{mansfield}:

\begin{equation}
\label{eq:fk} H = \sum_i V_{ss}(l_i) + \sum_jV_{sb}({\bf r}_j) + \Gamma N, 
\end{equation}

\noindent where $i$ runs over all NN bonds of length $l_i$ between
surface atoms, and $j$ runs over all atoms at positions ${\bf r}_j$ in
the surface layer. $N$ is the total number of atoms, and $\Gamma$ is a
chemical potential that contains information about the energy required
to incorporate an atom into the surface layer; this depends on where
the atom comes from (bulk, step edge, adatom, etc.)

To obtain $V_{ss}$, $V_{sb}$ and $\Gamma$, we have performed {\em ab
initio} density functional theory calculations to study the energetics
of compressing the surface layer, making surface stacking faults, and
extracting atoms from various sites. We use the PWSCF package
\cite{pwscf}, with a plane wave basis with a cut-off of 20 Ry,
ultrasoft pseudopotentials \cite{ultra}, and the local density
approximation to the exchange-correlation potential.  Surface
calculations have been carried out using supercells with 9 layers of
atoms, separated by a vacuum of 6-layer thickness. A Monkhorst-Pack
grid corresponding to 27 {\bf k}-points in the irreducible part of the
surface Brillouin zone is used to sample reciprocal space for
calculations using a $(1\times1)$ surface cell; the grid is varied
commensurately when using larger surface cells.  We obtain the bulk
lattice constant as $a_0=3.92 \AA$, and find that the first interlayer
spacing $d_{12}$ is slightly expanded by 0.3\% relative to the bulk
interlayer spacing, and the next two interlayer spacings $d_{23}$ and
$d_{34}$ are slightly contracted by 0.55\% and 0.18\% respectively. We
obtain the (unreconstructed) surface energy $\gamma$ and surface stress
$\sigma$ as 9.13 and 29.45 ${\rm mRy}/\AA^2$ respectively. These
results are in good agreement with experiments and previous
calculations \cite{needs,feibelman,boisvert}.

The trickiest part of the parametrization concerns $V_{ss}$, the
interaction between surface atoms. Earlier authors (who studied other
systems), either used ``physically reasonable" parameters
\cite{hamiltonagpt}, or did calculations to study the energetics of a
monolayer of atoms on a jellium \cite{takeuchi}; the ambiguity in the
latter approach arises from the uncertainty in the choice of jellium
density. We have chosen instead to calculate the variation in the
surface stress when compressing a slab of atoms \footnote{In principle,
only the top layer should be compressed; however for commensurate
compressions of 2/3, 3/4 and 4/5 we have verified that the two
procedures give essentially the same result.}. The distance $d_{12}$ is
allowed to relax, and it is therefore a good approximation to assume
that the variation in surface stress comes entirely from the
surface-surface bonds that we are interested in. We use these results
to parametrize a Morse potential $V_{ss}=A_0\{1-\exp[-A_1(a-b)]\}^2$;
we obtain $A_0$ = 60.2 mRy, $A_1$ = 2.062 $\AA^{-1}$, and $b$ = 2.638
$\AA$, i.e., surface bonds would like to shorten their length by 4.7\%
from the bulk NN distance $a=a_0/\sqrt 2=2.77 \AA$.

To obtain $V_{sb}$, we study surface stacking faults. We find that the
energy cost (per surface atom) of having the surface layer occupy HCP, top and bridge sites
instead of the most favored FCC site is 5.05, 11.85 and 5.89 mRy
respectively. These values are then used to expand $V_{sb}$ in a 2D
Fourier series, using the first two shells of reciprocal lattice
vectors for the 2D surface lattice \cite{takeuchi,narasimhan}. Fig.~1
shows $V_{sb}$ for a line cutting through the surface along the
$[11\bar2]$ direction. Note that though the HCP site is a local
minimum, it lies considerably above the FCC site in energy.

\begin{figure} 
\includegraphics[width=60mm]{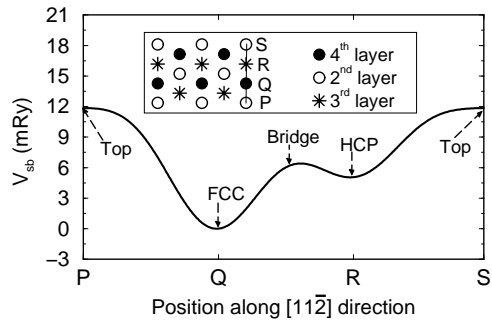}%
\caption{The
inset shows a top view of the Pt(111) surface, while the main graph
shows the surface-bulk potential $V_{sb}$ along the line PQRS marked in
the inset.} 
\label{F_fig1} 
\end{figure}

Finally, to obtain $\Gamma$, we use our {\em ab initio} results for the
bulk cohesive energy, the surface energy, the adatom adsorption energy
and the energy needed to detach an atom from a kink site at a step edge
\cite{unpub} to obtain $\Gamma_b$ = 50.56 mRy for bulk atoms,
$\Gamma_a$ = -67.12 mRy for adatoms,  and $\Gamma_k$ = 50.4 mRy for
kink atoms.

To see how stable the unreconstructed surface is, one can map the 2D
problem onto a 1D FK model \cite{mansfield}, and evaluate the
dimensionless parameter $R=(3\pi a)({4\over3} \sigma -
\gamma)/8\sqrt{2kW}$, where $k$ is the spring constant for NN surface
bonds and $W$ is the amplitude of $V_{sb}$ along the zigzag line
connecting adjacent FCC and HCP sites. If $R >/< 1$, the surface will /
will not reconstruct.  Mansfield and Needs \cite{mansfield} have
obtained $R=0.73$ for Pt(111) and $R=0.44$ for Au(111), i.e., they
predict that unreconstructed Au(111) should be more stable than
Pt(111), whereas it is well known that Au(111) reconstructs even at low
temperatures. We believe that the deficiency lies not with their model
but with  the values they used to obtain $R$.  Upon approximating
$V_{ss}$ by a harmonic potential in the region of interest, we obtain
$k$=370 mRy/$\AA^2$. Using $W$ = 5.89 mRy, we get $R$ = 0.998,
 i.e., we find that the unreconstructed surface is {\em only just}
stable. This explains why it is relatively easy to induce the surface
to reconstruct by increasing the temperature (reducing $W$) or by
placing the surface in a supersaturated vapor (effectively  increasing
the numerator of $R$).

For a more precise determination of structural stability, we work with
the full 2D Hamiltonian of Eq.~1.  We study the variation in the
surface energy $\gamma$ as a function of $\Delta\rho$, the excess
density in the surface layer (relative to the unreconstructed case),
and $\theta$, the angle between the unit vectors of the top layer and
the substrate.  (Though optimizing $\theta$ makes only a very small
difference to $\gamma$, we will show below that allowing for $\theta
\ne 0$ is crucial to explain the ``rotors" at intersections of domain
walls.) In all cases, we start with a uniformly compressed layer of
surface atoms, and obtain the atomic positions that minimize $\gamma$,
using a conjugate gradient algorithm.

\begin{figure} 
\includegraphics[width=75mm]{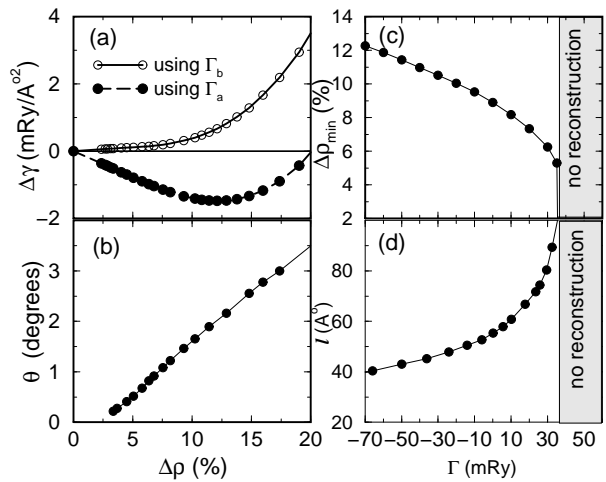}%
\caption{Results from the 2D Frenkel-Kontorova model for isotropic
reconstruction. (a) and (b) show how $\Delta\gamma$, the change in the
surface energy (relative to the unreconstructed surface) and $\theta$,
the angle of rotation of the top layer, depend on the excess density
$\Delta\rho$; (c) and (d) show how the optimal excess density
$\Delta\rho_{min}$ and the periodicity $l$ vary with the chemical
potential $\Gamma$. } 
\label{F_fig2} 
\end{figure}

For Pt(111), we find that isotropically compressed structures, with
threefold symmetry and sixfold atomic coordination everywhere, are
always lower in energy than the uniaxial ``stripe" patterns or
structures with point dislocations.  Fig. 2(a) shows how $\Delta\gamma$,
the difference in $\gamma$ for the reconstructed and unreconstructed
surfaces, varies with $\Delta\rho$. Each point is also optimized with
respect to $\theta$; Fig. 2(b) shows that $\theta$ varies approximately
linearly with $\Delta\rho$ and is very small.  If the surface
reconstructs, $\Delta \gamma$ will have a minimum at a non-zero value
$\Delta\rho_{min}$. Under normal conditions, extra atoms are obtained
from the bulk or step edges, and $\Gamma \simeq \Gamma_b$. It is clear
from the figure that under these conditions, the surface will not
reconstruct, which agrees with experiment. However, if adatoms are
available for incorporation in the surface layer, $\Gamma$ is lowered,
up to a minimum of $\Gamma_a$, and the surface then reconstructs.
Fig.~2(c) shows how $\Delta\rho_{min}$ varies as $\Gamma$ varies
between $\Gamma_a$ and $\Gamma_b$. The surface reconstructs only  if
$\Gamma < $ 36 mRy.  As $\Gamma$ decreases and $\Delta\rho$ increases,
the periodicity $l$ of the reconstruction decreases; Fig 2(d) shows how
$l$ varies with $\Gamma$.

Our most striking results are obtained upon examining the domain
patterns using a simple technique to obtain the surface corrugation and
thus simulate STM images. The height of each atom in the unit cell is
obtained using a 2D Fourier expansion similar to that used for
$V_{sb}$, expanding about the heights at FCC, HCP, bridge and top
sites. From our {\em ab initio} calculations on surface stacking
faults, we obtain these heights as 0, 0.03, 0.29 and 0.06 $\AA$
respectively. These are smaller than the corrugations measured
experimentally \cite{bott}. However, we find that {\em ab initio}
simulations of the reconstructed structure (for small cell sizes
accessible to computation) support our smaller values \cite{unpub}. As
has been reported recently for Au(111), it appears that the STM
exaggerates the value of the corrugation \cite{stm}; in any event, we
are primarily interested in the domain patterns, rather than in the
absolute heights.

\begin{figure} 
\includegraphics[width=65mm]{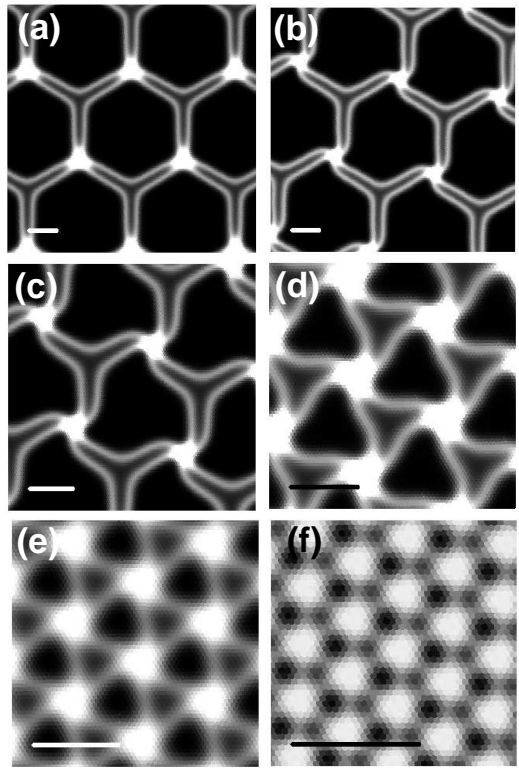}%
\caption{
Simulated STM images of the Pt(111) surface, as $\Delta\rho$ is varied
from 2.9\% to 21\%. Individual atoms are shaded according to their
height; lighter atoms are higher. In (a) the top layer is aligned with
the substrate, in (b) to (f), the angle betwen the top layer and
substrate has been optimized. (a) and (b) are the honeycomb, (c) and
(d) are threefold whorls (e) is the bright star, and (f) is the Moir\'e
pattern. The black/white line in each image has a length of 50 $\AA$}
\label{F_fig3} 
\end{figure}

A sequence of  simulated STM images is shown in Fig.~3.  In all six
images, black, dark grey, light grey and white areas correspond to
regions where atoms sit at FCC, HCP, bridge and top sites respectively.
A non-linear grey scale has been used to show the domain walls (bridge
sites) clearly.  Fig.~3(a) shows the pattern obtained when
$\Delta\rho$=2.9\%, and $\theta=0$. Large hexagonal FCC domains are
separated from narrow HCP domains by domain walls that form a honeycomb
network. Alternate vertices of the hexagons are extremely bright; atoms
here sit at top sites.  Between these bright vertices, there is a
``three-pointed star" similar to that obtained in an earlier simulation
of the nucleation of the reconstruction \cite{jacobsen}.
 Fig.~3(b) shows the structure obtained for the same $\Delta\rho$, but
with $\theta$ at its optimal value of $0.14^{\circ}$.  It is evident
from a comparison of the two images that breaking the
symmetry by even this tiny amount has a big impact on the
structure. The bright vertices are now transformed into ``bright
rotors", which are indeed observed experimentally \cite{hohage}.  The
similarity between Fig~3(b) and the experimental STM images
\cite{bott,hohage} is striking.  Our finding that the rotors arise from
a rotation of the top layer relative to the substrate explains why, in
a given region, experiments tend to find either all clockwise or all
anticlockwise rotors. The honeycomb pattern occurs in a region where
$l$ is very sensitive to $\Gamma$; we believe this explains why STM
images usually show irregular hexagons.

As $\Gamma$ is lowered, the surface layer densifies further. Figs.~3
(c) and (d) show the structures obtained when $\Delta\rho$ is increased
to 4.0\% and 6.8\% respectively.  The honeycomb transforms into a
pattern of interlocking wavy triangles. Such structures have been
observed on dense terraces of Pt(111) \cite{hohage}.  This structure
has also been seen for Na/Au(111) \cite{naau} and for 3 ML of
Cu/Ru(0001) \cite{curu}.  As $\Gamma$ is lowered and $\Delta\rho$
increased still further, the wavy domain walls straighten out,
resulting in the ``bright star" pattern shown in Fig.~3(e), for a value
of $\Delta\rho$= 10.2\%. This pattern has not been seen on Pt(111) as
it requires a very low $\Gamma$ (high adatom coverage); however it is
seen on other surfaces, e.g., Ni/Ru(0001) \cite{niru}.
$\Delta\rho_{min}$ for Pt(111) has a maximum value of 12\% (when
$\Gamma=\Gamma_a$); it is nevertheless instructive to see what happens
for larger $\Delta\rho$.  The triangular domains of the bright star
transform into hexagons, resulting in the Moir\'e pattern shown in
Fig~3(f). The density distribution on the surface is now almost
uniform. This pattern is also observed experimentally, e.g., for 4 ML
of Cu/Ru(0001) \cite{curu}.

The sequence of images shown in Fig.~3 makes it clear that all these
structures ``morph" smoothly into one another and are topologically
equivalent. Progressively larger values of $\Delta\rho$ can be favored
either by lowering $\Gamma$ [as for Pt(111)], or by increasing the
surface stress -- e.g., by depositing more overlayers in a
heteroepitaxial system. Indeed, a similar progression of structures
with overlayer thickness has been observed for the Cu/Ru(0001) system
\cite{curu,hamiltoncuru}. Our results show that the same physics is
operating in all the systems mentioned above, and it is only slight
changes in parameters that are responsible for the various
domain patterns observed on different surfaces.

To summarize: we have shown that the unreconstructed Pt(111) surface is
only just stable, and can thus be easily induced to reconstruct.  We
have confirmed that Pt(111) will reconstruct if the chemical potential
is lowered by the presence of a large number of adatoms.  In addition
to the methods tried to date, we suggest that it should also be
possible to induce the surface to reconstruct in an electrochemical
environment or by depositing alkali metals on it. The domain patterns
we obtain for Pt(111) are in excellent agreement with experiment. We
have also shown that slight variations in conditions can lead to many
of the other domain patterns seen on similar surfaces. We have shown
that the periodicity of the reconstruction can be controlled by varying
the chemical potential; this is important if one would
like to use such surfaces as templates for growing ordered
nanostructures.

We thank S. de Gironcoli for providing the Pt ultrasoft
pseudopotential, and X. Gonze, I.K. Robinson, H. Brune and U.V.
Waghmare for helpful discussions.



\end{document}